# Rivals for the crown: Reply to Opthof and Leydesdorff

Anthony F.J. van Raan, Thed N. van Leeuwen, Martijn S. Visser,
Nees Jan van Eck, and Ludo Waltman

Centre for Science and Technology Studies, Leiden University, The Netherlands
{vanraan, leeuwen, visser, ecknjpvan, waltmanlr}@cwts.leidenuniv.nl

We reply to the criticism of Opthof and Leydesdorff on the way in which our institute applies journal and field normalizations to citation counts. We point out why we believe most of the criticism is unjustified, but we also indicate where we think Opthof and Leydesdorff raise a valid point.

## 1. Introduction

Opthof and Leydesdorff (in press; henceforth O&L) criticize the way in which our institute, the Centre for Science and Technology Studies (CWTS) of Leiden University, applies journal and field normalizations to citation counts. The criticism of O&L focuses on two of our citation-based indicators of research performance. These indicators are our so-called crown indicator, which normalizes for differences among fields, and a related indicator that normalizes for differences among journals. To illustrate their criticism, O&L use a research performance evaluation of the Academic Medical Center (AMC) of the University of Amsterdam. In this reply, we will comment on the various issues raised by O&L. We will point out why we believe most of their criticism is unjustified. We will also indicate where we think O&L raise a valid point.

## 2. Inaccuracies and omissions in the paper by O&L

Before replying to the main issues raised by O&L, we first would like to take the opportunity to point out some important inaccuracies and omissions in the paper by O&L. O&L refer to a confidential CWTS report in which an evaluation of the research performance of the AMC is presented. However, such a report does not exist. What O&L refer to is actually a report of the AMC itself in which it evaluates its own research performance (AMC, 2008). This report is based on data that CWTS has provided, but the report has been produced by AMC staff, not by CWTS staff. In fact, until very recently CWTS was not even aware of the existence of this report. Obviously, CWTS cannot take responsibility for a report that it has never seen. What CWTS does take responsibility for is the data that it has provided to the AMC.

It should further be pointed out that O&L cite the AMC report incorrectly and selectively. According to O&L, the report states that a citation score of 0.80 (relative to the world average of 1.00) represents the "borderline value of underperformance". However, the report does not make any such statement. Instead, the report classifies a citation score of 0.80 as "below the world average" (AMC, 2008, p. 8, translated from Dutch). Moreover, the report makes the cautionary comment that "in case of a low citation impact score the conclusion of underperformance is not directly justified" (AMC, 2008, p. 88, translated from Dutch). This clearly contradicts the way in which O&L cite the report.

CWTS calculates the normalized citation score of a set of publications as the ratio of the average observed number of citations of the publications and the average expected number of citations of the publications (e.g., Moed, De Bruin, & Van Leeuwen, 1995; Van Raan, 2005). The expected number of citations of a publication is determined by the field or the journal in which the publication has been published, the age of the publication, and the document type of the publication (i.e., article, letter, or review). O&L argue that this method for normalizing citation counts should not be used. Instead, an observed/expected ratio should be calculated for each publication separately and the average of the ratios calculated for all publications should be used as a normalized citation score. O&L do not acknowledge that this proposal is not new. The same proposal was made earlier by Lundberg (2007), whose work is not mentioned by O&L. In fact, the alternative normalization method proposed by O&L is already being used by various bibliometric institutes and research groups around the world (e.g., Rehn & Kronman, 2008; SCImago Research Group, 2009).

## 3. Reply to the main arguments of O&L

O&L put forward three main arguments in favor of the alternative normalization method:

1. The normalization method of CWTS implies “a violation of the order of operations which prescribes that divisions precede additions”. The alternative normalization method does not have this problem.
2. The alternative normalization method has the advantage that it yields normally distributed variables and, consequently, that it allows one “to test for the significance of the deviation of the test set from the reference set”.
3. The normalization method of CWTS “assumes that more highly cited papers should carry more weight in the index”, while in fact “all papers should have an equal weight in an index”. The alternative normalization method indeed weighs all publications equally.

We will reply to each of these arguments in turn.

The first argument completely misses the point. The order of operations, which states that multiplication and division precede addition and subtraction, is nothing more than a convention that indicates how mathematical expressions are to be interpreted. The order of operations is not meant to be prescriptive and hence does not indicate that multiplication and division *should* be performed before addition and subtraction. Because of this, the order of operations argument is irrelevant in the choice between the two normalization methods.

The second argument of O&L is also not very relevant. It is true that the alternative normalization method allows one to perform statistical significance tests and to construct confidence intervals. However, the same can be done when the normalization method of CWTS is used. In a large number of studies conducted by CWTS, the significance test of Schubert and Glänzel (1983; see also Moed et al., 1995) has for example been employed. An alternative strategy could be the use of bootstrapping techniques (e.g., Efron & Tibshirani, 1993; Spiegelhalter & Goldstein, 2009).

This brings us to the third argument of O&L. This is a much more interesting argument. (Note that the argument is not new. The same argument was put forward by Lundberg (2007).) According to O&L, the normalization method of CWTS “assumes that more highly cited papers should carry more weight in the index”. This is not entirely correct. In our normalization method, the weight that is given to a publication depends on the expected number of citations of the publication, not on the publication's observed number of citations (Lundberg, 2007; Waltman, Van Eck, Van Leeuwen, Visser, & Van Raan, 2010). It is true, however, that our normalization method does not weigh all publications equally. Publications from fields or journals with a high expected number of citations have more weight than publications from fields or journals with a low expected number of citations. Similarly, older publications have more weight than newer ones and ordinary articles generally have more weight than letters. O&L argue that "all papers should have an equal weight in an index". We believe that this is too simple. Writing a letter generally takes less time and effort than writing an ordinary article. It therefore makes sense to weigh these two types of publications differently. Very recent publications have not had much time to earn citations, and their citation impact therefore cannot be determined accurately. This could be a reason to weigh older and newer publications differently. In the case of publications from different fields, however, we think O&L have a valid point. In general, there does not seem to be a good reason to weigh publications from different fields differently. For some time already, we have been thinking at CWTS about revising our crown indicator in such a way that publications from different fields are weighed equally. This means that for the purpose of normalizing for field differences we need to switch to the alternative normalization method discussed by O&L and earlier by Lundberg (2007). In a paper that we have just finished (Waltman et al., 2010), we study this issue in detail and we provide a number of arguments why for field normalization purposes the alternative normalization method is indeed preferable over our current method. Based on the various arguments, we plan to adopt the alternative normalization method in future performance evaluation studies. We emphasize, however, that things are more complicated when it comes to normalizing for differences among document types and for differences among publications of different ages. As discussed above, for differences among document types the alternative normalization method seems less appropriate than our current method. For differences among publications of different ages, the alternative normalization method requires special care, since the accuracy of performance indicators may be reduced due to the effect of very recent publications. An empirical illustration of this issue will be given in the next section.

## 4. Empirical analysis

In order to evaluate the practical differences between our current normalization method and the alternative method, we calculated field normalized citation scores according to both methods. Results are presented for the researchers for whom CWTS provided performance indicators to the AMC. O&L illustrate their argument with empirical results for 7 AMC researchers. As regards their analysis, the following comments are in order:

- O&L use journal normalized citation scores rather than field normalized citation scores. In the case of field normalization, we agree with O&L that the alternative normalization method is preferable over our current method. However, in the case of journal normalization, we believe that things are less clear-cut. On the one hand the idea of weighing all publications equally (as the alternative normalization method does) may seem appealing, but on the other hand one could also argue that publications in high impact journals should have more weight than publications in low impact journals. Because the alternative normalization method seems less debatable in the case of field normalization than in the case of journal normalization, we focus on field normalization in our analysis.

- O&L wrongly suggest that discrepancies between the number of publications retrieved by them and by CWTS may be due to CWTS leaving out review articles. The AMC report clearly states that review articles have been included in the analysis (AMC 2008, p. 6). On the same page, the report also makes clear that the publication data used in the analysis have been collected by the AMC itself and have been checked internally.
- The citation counts used by O&L were recorded at a much later moment (October 2009) than the citation counts used by CWTS. The analysis of CWTS reports citation counts accumulated by the end of 2006. This means that in the analysis of O&L citation counts for recent publications are much more robust.

It is further important to realize that in the case of the performance evaluation of the AMC, performance indicators were calculated at the level of individual researchers. The publications of an individual researcher tend to be categorized into a limited number of fields. Because of this, the way in which field normalization is performed will generally have a relatively small effect in the case of individual researchers. Most likely, the effect of the way in which normalization for publication age is performed will be much more substantial.

Figure 1 compares the two normalization methods for 204 AMC researchers with 20 or more publications indexed by Web of Science.[1] On the horizontal axis, citation scores are given as provided by CWTS to the AMC. These scores were calculated according to our current normalization method. The vertical axis indicates citation scores calculated according to the alternative normalization method. Contrary to what is suggested by O&L, the data show no clear evidence of a systematic underrating of low-ranked researchers.

[1] CWTS provided data for 257 researchers to the AMC. In the AMC report, data is presented for 256 researchers. O&L mention 232 researchers, but it is not clear to us how exactly O&L made their selection of researchers.

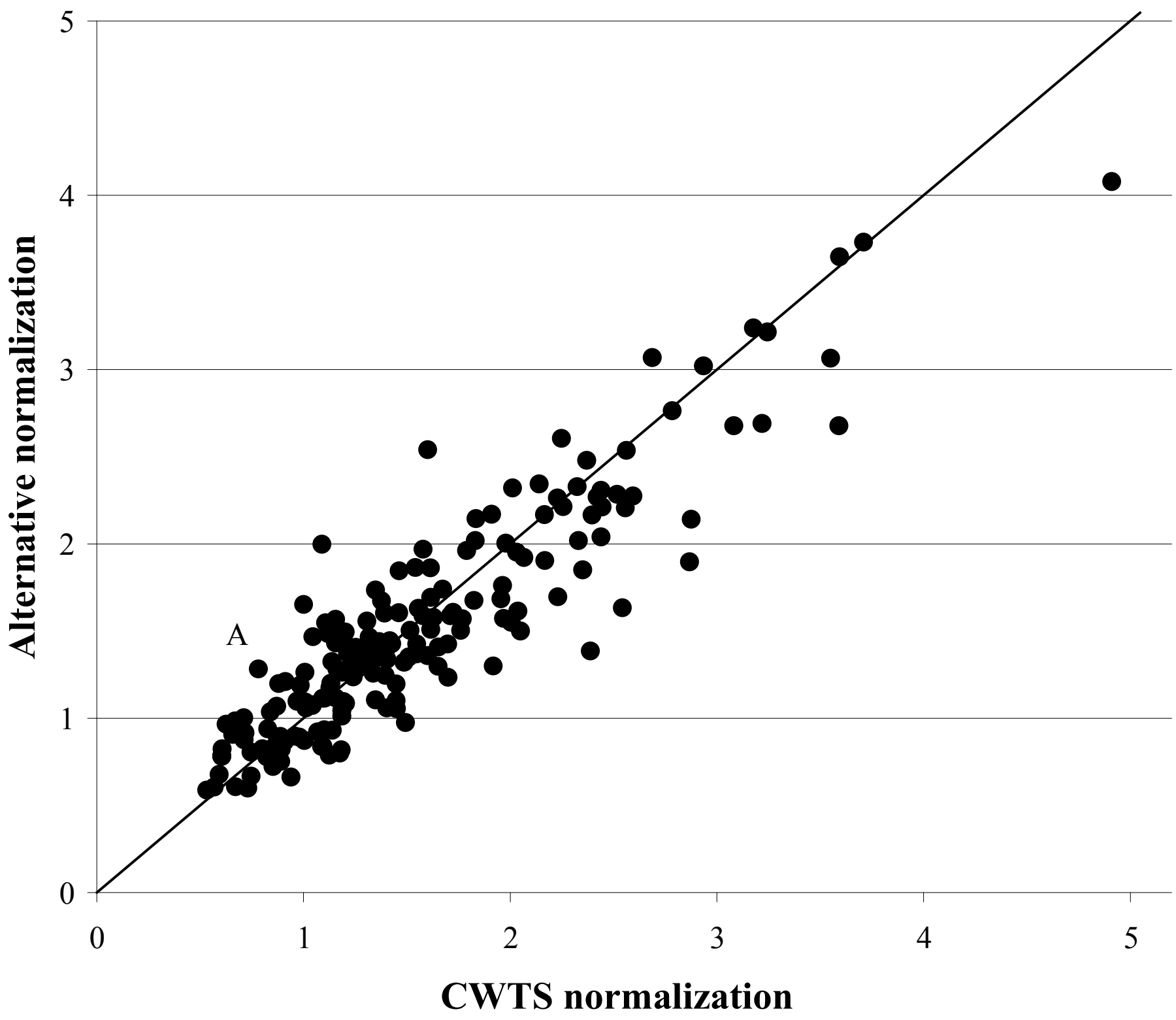

Figure 1. Field normalized citation scores for AMC researchers with 20 or more Web of Science publications between 1997 and 2006 (citations counted up to 2006).

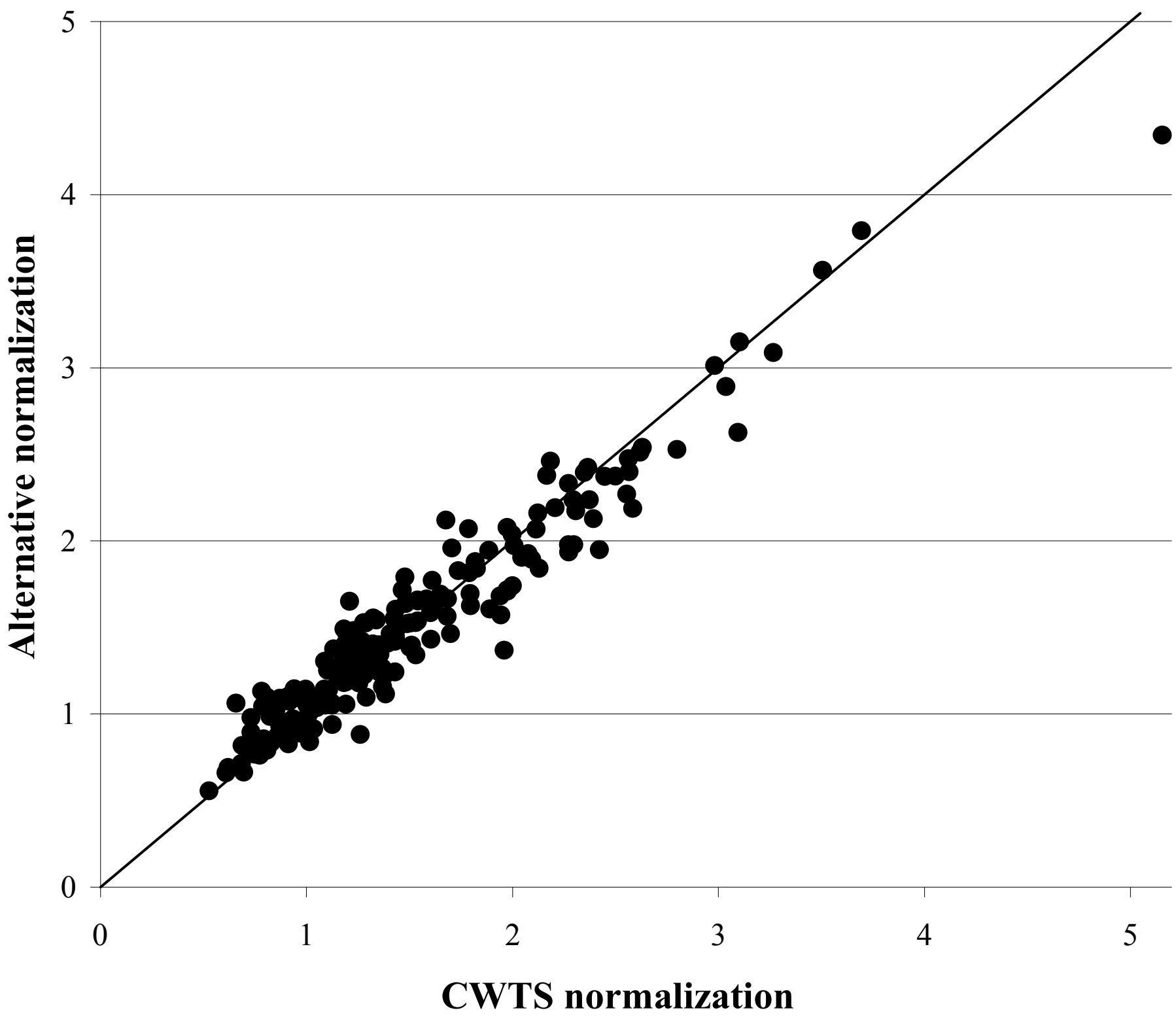

Figure 2. Field normalized citation scores for AMC researchers with 20 or more Web of Science publications between 1997 and 2006 (citations counted up to 2008).

Of particular interest are those researchers for whom the difference between the two normalization methods leads to a significantly different interpretation of their citation impact. As an example we look at researcher A shown in Figure 1. This researcher has 53 publications and 349 citations. According to the alternative normalization method, the citation impact of researcher A is well above average (1.28). In contrast, our current normalization method indicates a below average citation impact (0.78). However, the difference between these two outcomes can largely be attributed to 10 citations received by 3 publications published in 2006. Because of low expected citation scores for very recent publications (citations were counted up to the end of 2006), the citation scores of these 3 publications exceed expectation on average more than 8 times.

Extremely high normalized citation scores for very recent publications can be observed for many researchers located far above the diagonal in Figure 1. To a large extent, the rankings of these researchers should be qualified as spurious since they are based on very low expected citation scores. In fact, many outliers in Figure 1 disappear when the length of the citation window is extended by two years, as is depicted in Figure 2. For the greater part, this should be attributed to differences in the outcomes of the alternative normalization method as our current normalization method is less sensitive to an extension of the citation window.

The above empirical analysis does not disqualify the alternative normalization method but merely demonstrates that the method has an important practical drawback. The alternative normalization method may become unstable in the case of very low expected citation scores. This issue should be dealt with carefully but in itself does not impede the use of the alternative normalization method.

## 5. Use of Web of Science subject categories

Finally, let us reply to another issue raised by O&L. To normalize citation counts for field differences, one typically uses a classification scheme for assigning publications to fields. CWTS uses Web of Science (WoS) subject categories for this purpose. O&L criticize the use of WoS subject categories because these categories “sometimes heavily overlap and are often misguided”. We do not see why overlap of categories should be considered problematic. Overlap of categories may simply reflect the fuzziness of disciplinary boundaries and the multidisciplinary character of many journals. We also do not agree with O&L that WoS subject categories are often misguided. Although some categories may not be sufficiently homogeneous, as suggested by Boyack, Klavans, and Börner (2005), we are not aware of any convincing evidence of large-scale inaccuracies in the classification scheme of WoS.

As an alternative to WoS subject categories, O&L suggest the use of disciplinary classification schemes, for example based on the Chemical Abstracts database or the Medical Subject Headings of the MEDLINE database. In certain cases, this may be a useful approach, and in fact CWTS is also experimenting with this approach (Van Leeuwen & Calero Medina, 2009). However, for research performance assessment at higher aggregation levels, for example at the level of countries, universities, or other institutes with a broad scope, disciplinary classification schemes do not offer a solution. This is because many disciplines do not have their own classification scheme and also because the combined use of several different (possibly overlapping) classification schemes is impractical for various reasons.

Research into the most appropriate way of delineating fields for normalization purposes (e.g., Adams, Gurney, & Jackson, 2008; Glänzel, Thijs, Schubert, &

Debackere, 2009; Van Leeuwen & Calero Medina, 2009; Zitt, Ramanana-Rahary, & Bassecoulard, 2005) is important, and more research into this issue is certainly needed. However, since disciplinary boundaries are intrinsically fuzzy, any classification scheme involves some arbitrariness and a completely satisfactory scheme simply does not exist. An interesting alternative approach therefore is to try to normalize for field differences without using a classification scheme (Moed, in press; Zitt & Small, 2008). At CWTS, we are currently investigating the general applicability of such a source-normalized approach (Moed, in press) to research performance assessment.

## 6. Conclusion

An open scientific debate on the advantages and disadvantages of different citation-based indicators of research performance is crucial for bibliometric performance assessment to be conducted in the most proper way. At CWTS, we therefore very much welcome constructive criticism on our approach to research performance assessment. As we have pointed out, the criticism of O&L is inaccurate in various respects, in particular in the way in which it treats the AMC report (AMC, 2008). In addition, some of the main arguments of O&L are seriously flawed. Rather than providing new insights, these arguments create unnecessary confusion. Having said this, we acknowledge that O&L also raise a valid and important point. The normalization method of the CWTS crown indicator has the unsatisfactory property that it gives more weight to publications from fields with a high expected number of citations than to publications from fields with a low expected number of citations. This point was also made by Lundberg (2007), and the issue is studied in detail in a paper that we have just finished (Waltman et al., 2010). At CWTS, we are currently revising our crown indicator in such a way that publications from different fields are weighed equally.

We very much agree with O&L that citation-based performance indicators should be used carefully, especially at lower levels of aggregation, such as at the level of individual researchers or small research groups. At CWTS, we always communicate this in a clear and open manner to our customers. In the case of the research performance evaluation of the AMC, the data provided by CWTS indeed seem to have been interpreted with great care (e.g., AMC, 2008, p. 88). O&L also argue that "the transparency and traceability of (bibliometric) indicators should be one of the primary objectives". This point is fully shared by CWTS. Contrary to what O&L suggest, customers of CWTS can always get access to the raw data based on which CWTS has calculated its performance indicators. We consider this essential for proper bibliometric performance assessment.

## Acknowledgment

We are grateful to the AMC for providing us with a copy of its internal and confidential performance evaluation report (AMC, 2008) and for allowing us to use this report in our reply to O&L.